\documentclass[review=false]{acmart}
\usepackage{graphicx} 
\usepackage{subcaption} 
\AtBeginDocument{%
  }

\setcopyright{acmlicensed}
\copyrightyear{2018}
\acmYear{2018}
\acmDOI{XXXXXXX.XXXXXXX}

\acmConference[Conference acronym 'XX]{Make sure to enter the correct
  conference title from your rights confirmation emai}{June 03--05,
  2018}{Woodstock, NY}
\acmISBN{978-1-4503-XXXX-X/18/06}

\begin{document}

\title[Social Media Informatics for Sustainable Cities and Societies]{Social Media Informatics for Sustainable Cities and Societies: An Overview of the
Applications, associated Challenges, and Potential Solutions}

\author{Jebran Khan}
\email{jebran@ajou.ac.kr}
\affiliation{%
  \institution{Department of Artificial Intelligence, Ajou University}
  \city{Suwon}
  \country{South Korea}
}

\author{Kashif Ahmad}
\email{kashif.ahmad@mtu.ie}
\affiliation{%
  \institution{Munster Technological University}
  \city{Cork}
  \country{Ireland}}

\author{Senthil Kumar Jagatheesaperumal}
\email{senthilkumarj@mepcoeng.ac.in}
\affiliation{%
  \institution{Department of Electronics \& Communication Engineering, Mepco Schlenk Engineering College}
  \city{Sivakasi}
  \country{India}
}

\author{Nasir Ahmad}
\email{nasir.ahmad@mtu.ie}
\affiliation{%
 \institution{Munster Technological University}
 \city{Cork}
 \country{Ireland}}

\author{Kyung-Ah Sohn}
\affiliation{%
  \institution{Department of Software and Computer Engineering, Ajou University}
  \city{Suwon}
  \country{South Korea},
  \institution{Department of Artificial Intelligence, Ajou University}
  \city{Suwon}
  \country{South Korea}
  }
 \email{kasohn@ajou.ac.kr}



\renewcommand{\shortauthors}{Khan et al.}

\begin{abstract}
  In the modern world, our cities and societies face several technological and societal challenges, such as rapid urbanization, global warming \& climate change, the digital divide, and social inequalities, increasing the need for more sustainable cities and societies. Addressing these challenges requires a multifaceted approach involving all the stakeholders, sustainable planning, efficient resource management, innovative solutions, and modern technologies. Like other modern technologies, social media informatics also plays its part in developing more sustainable and resilient cities and societies. Despite its limitations, social media informatics has proven very effective in various sustainable cities and society applications. In this paper, we review and analyze the role of social media informatics in sustainable cities and society by providing a detailed overview of its applications, associated challenges, and potential solutions. 
  This work is expected to provide a baseline for future research in the domain.
\end{abstract}


\received{20 February 2007}
\received[revised]{12 March 2009}
\received[accepted]{5 June 2009}

\maketitle

\section{Introduction}
\label{sec:introduction}
Recently, enormous growth has been noticed in the urban population, currently, globally, over 50\% of the world's population resides in urban areas \footnote{Urban Development: https://www.worldbank.org/en/topic/urbandevelopment/overview}. The numbers are expected to reach 68\% by 2050 \cite{ahmad2022developing}. Several socioeconomic and political factors contribute to this rapid urbanization. For instance, large cities offer more employment and educational opportunities and social and cultural attractions. However, it also has a significant environmental and social impact, and the concerned public authorities are urged to act quickly by maintaining a balance between the rapid expansion and resources and infrastructure of the cities \cite{ahmad2022developing}. This balance between resources and expansion of the cities requires a multifaceted approach involving sustainable planning, efficient resource management, and innovative solutions. This will lead to more sustainable cities and societies. 

The concept of sustainable cities has been around for some time, however, the term sustainable societies is relatively new. Sustainable cities are generally associated with the environmentally friendly development of cities by promoting sustainable production and consumption patterns. However, the term sustainable society extends the concept of sustainable cities by bringing several important components, such as inclusive governance, social equity, economic viability, and cultural vibrancy. Sustainable cities and societies strive for a fine balance among environmental, societal, and economic aspects of urban life, ensuring the long-term feasibility of quality of life for residents through proper planning and transformation via modern technologies. Environmental sustainability, which prioritizes and advocates a responsible interaction with the environment, involves several elements, such as promoting renewable sources of energy (hydro, wind, and solar), waste management (e.g., recycling or proper disposal), afforestation (e.g., planting more trees and halting deforestation), and sustainable transportation (e.g., walking, cycling, and encouraging zero emission). Social Sustainability, on the other hand, involves inclusive governance (e.g., involving or considering all groups of citizens in policy-making), equal opportunities and public services (e.g., education, healthcare, jobs, and transportation for all), and justice. Similarly, economic Sustainability focuses on the creation of job opportunities and supporting local economic growth, etc.,    

Considering the importance and benefits of sustainable cities and societies, the United Nations (UN) has also initiated several programs supporting sustainable cities and societies in different ways. For instance,  in 2015, the UN General Assembly defined 17 goals for sustainable development, which are known as ''Sustainable Development Goals (SDGs)''. The majority of these goals aim at inclusiveness, safety, resilience, and sustainability of cities and societies. Moreover, the UN has a dedicated working group/agency namely ''UN-Habitat'' advocating for socially and environmentally sustainable towns and cities. 

Despite the significant efforts, several economic, social, environmental, technical, and political challenges associated with developing sustainable cities and societies are yet to be resolved. For instance, sustainability initiatives/projects require an initial investment for the new infrastructure or upgrading the existing infrastructure to support sustainability initiatives. Moreover, generally, the public tends to be reluctant to such initiatives due to cultural resistance, which requires public awareness \cite{puiu2021recycling}. Moreover, addressing the gap among different socioeconomic groups and the disparities in access to sustainable resources and services, such as healthcare, education, and transportation, is also very challenging. 

To overcome these challenges, a multifaceted approach is needed by involving all the stakeholders including public authorities, the business community, and the general public and communities. Modern technologies, such as Artificial Intelligence (AI), Machine Learning (ML), Data Analytics, the Internet of Things (IoTs), material sciences, Building Management Systems (BMS), Sensors, and telecommunication technologies are playing their part in providing innovative solutions to address the diverse challenges of sustainable cities and societies, such as resource management, environmental conservation, and public awareness \cite{ahad2020enabling}. Like other modern technologies, social media informatics also plays a crucial role in developing sustainable cities and societies by leveraging the data and information shared on different platforms, such as Twitter, Facebook, and Instagram, for different tasks and applications to enhance urban living and foster sustainable practices. For instance, it could facilitate inclusive decision-making by engaging them at a larger scale and identifying community needs \cite{auyb2024social}. Similarly, social media could be used as a medium of communication for sharing real-time information and promoting sustainable practices \cite{chalmeta2024framework}.    

Social media informatics has been proven very effective in a diversified set of smart city applications, such as disaster management \cite{alam2023role}, feedback on public services \cite{ahmad2022social}, real-time information on roads during floods \cite{ahmad2019automatic}, and sustainability transition in tourism and agriculture \cite{hysa2021social}, leading to more sustainable and effective solutions. However, several challenges are associated with extracting meaningful insights from social media outlets \cite{said2019natural}. Generally, social media informatics involves several steps to extract meaningful insights from the raw data, such as collecting data from different platforms using the corresponding APIs, pre-processing (e.g., data cleaning), data annotation (for supervised classification), analysis, and data visualization. Each of these steps has its challenges. For instance, data access restrictions and API limitations are the biggest challenges in data collection. Additionally, social media data is generally unstructured and inconsistent, which makes it difficult to process. Social media text is subject to several intentional and unintentional textual variations, posing several challenges in pre-processing. Moreover,  processing multi-modal social media data is also associated with several challenges. In this paper, we explore the potential of social media informatics in sustainable cities and societies by discussing the potential applications, associated challenges, and potential solutions for making social media data more effective in building sustainable cities and societies. 


\subsection{Scope of the Survey}
The survey revolves around social media informatics applications for sustainable cities and societies. The paper provides a detailed overview of various sustainable cities and societal applications of social media informatics. The paper particularly focuses on the key technical and societal challenges associated with social media and how social media informatics could help overcome these challenges. The paper also provides a detailed bibliometric analysis of the domain.

\subsection{Related Survey}
\label{sec:related_work}
Sustainable cities and societies have been among the most attractive areas for researchers over the last few years. During this time, several interesting papers including survey articles exploring different aspects of sustainable cities and societies have been published. Being a multidisciplinary research area, these surveys focus on certain aspects of sustainable cities and societies. Some surveys discuss the literature on general frameworks, models, and approaches to sustainable cities and societies \cite{karal2024systematic}. For instance, Bibri et al. \cite{bibri2017smart} provide an overview of various sustainable city models and smart city approaches reported in the literature. The authors discussed the strengths and weaknesses of the proposed models by providing a detailed comparative analysis of the models. Similarly, \cite{sodiq2019towards} provided a detailed overview of the sustainability principles and trends in the literature. Talwar et al. \cite{talwar2023systematic} focused on the environmental aspects of sustainable cities and societies, providing a systematic survey of photovoltaic-green roof systems proposed in the literature for smart and sustainable cities.  Yigitcanlar et al. \cite{yigitcanlar2015ecological} surveyed the literature on the ecological approaches in planning for sustainable cities. 

Several surveys also explored the technological aspects of sustainable cities and societies. For instance, Biasin et al. \cite{biasin2024blockchain}, Rejeb et al. \cite{rejeb2021blockchain}, and Siddiquee et al. \cite{siddiquee2022blockchain} surveyed the literature on blockchain technology for sustainable cities. Similarly, Alaeddini et al. \cite{alaeddini2023bibliometric} provided a bibliometric analysis of the literature on the convergence of AI and Blockchain for Smart and Sustainable cities. Fadhel et al. \cite{fadhel2024comprehensive} provided a systematic survey of Internet of Things (IoTs) data and information fusion methods for smart and sustainable cities. Similarly, Sun et al. \cite{sun2017big} provided a detailed overview of big data and geospatial information for sustainable cities. More recently, Balakrishnan et al. \cite{balakrishnan2023sustainable} provided a systematic survey on the use of social media platforms for the resilience of sustainable smart cities with a particular focus on COVID-19 case studies. The authors examined 12 papers on social media and other technologies for contributions to community resilience during COVID-19. 

In contrast to existing surveys, this paper provides a comprehensive survey of social media informatics for sustainable cities and societies by providing a detailed overview of several interesting applications. This survey also covers the literature on several key technical challenges and societal concerns associated with social media informatics
for sustainable cities and societies. The paper also discusses the potential solutions to these challenges and explores how social media informatics could be made more effective for sustainable cities and societies. We also provide a detailed bibliometric analysis of the topic to better highlight the trends in the domain. 

\subsection{Contributions}

This paper provides a detailed overview of social media informatics applications for sustainable cities and societies. The paper mainly focuses on the key applications, associated societal and technical challenges, and potential solutions. 

The main contributions of the paper are summarized as follows:

\begin{itemize}
    \item We examine the positive role played by social media informatics toward sustainable cities and societies by providing a detailed overview of its various applications for sustainable cities and societies. 
    \item The paper explores the literature on the major societal concerns and technical challenges to social media informatics in various applications, including generic and application-specific challenges.
    \item The paper also discusses potential solutions to these challenges, highlighting how a multifaceted approach involving all stakeholders could improve the effectiveness of social media informatics in developing sustainable cities and societies. 
\end{itemize}

The rest of this paper is organized as follows: Section~\ref{sec:applications} provides an overview of key applications of social media informatics in sustainable cities and societies. 
Section~\ref{sec:Challenges and Potential Solutions} discusses critical societal concerns and technical challenges in deploying social media informatics in these applications and the potential solutions. Section~\ref{sec:insights} summarizes insights and key lessons learned during this work. Finally, Section~\ref{sec:conclusion} concludes the paper.

\section{Applications of Social Media Analytics in Sustainable Cities and Societies}
\label{sec:applications}
\begin{figure*}[!t]
\centering
\centerline{\includegraphics[width=0.8\textwidth]{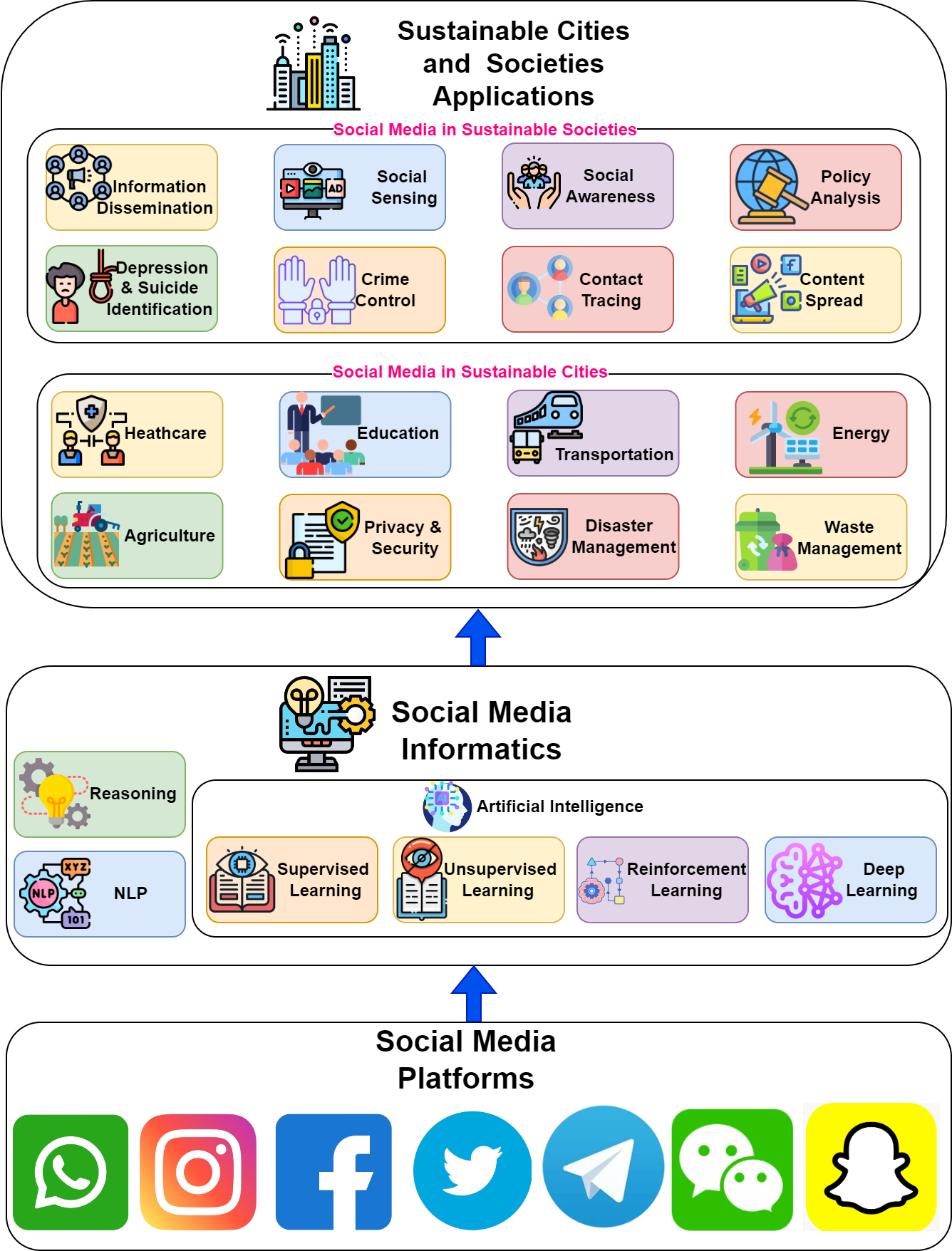}}
\caption{An overview of how social media informatics contributes to sustainable cities and societies. In the process, data from different social media platforms is collected; The data is then analyzed with different AI and data analytics techniques to extract meaningful insights for different applications.}
\label{fig:sustainablecitiesandsocities}
\end{figure*}

The power of social media analytics paves the path for efficient resource allocation and community engagement in pursuing sustainable urban development. This section focuses on the broader context of establishing sustainable cities and societies, considering the social media insights that facilitate inclusive dialogue and collective action towards shared environmental and social goals. Figure \ref{fig:sustainablecitiesandsocities} provides an overview of how social media informatics contributes to sustainable cities and societies. 
The next subsections provide a detailed overview of social media informatics's applications in sustainable cities and societies.

\subsection{Social Media in Sustainable Cities }

\begin{figure*}[!t]
\centering
\centerline{\includegraphics[width=1\textwidth]{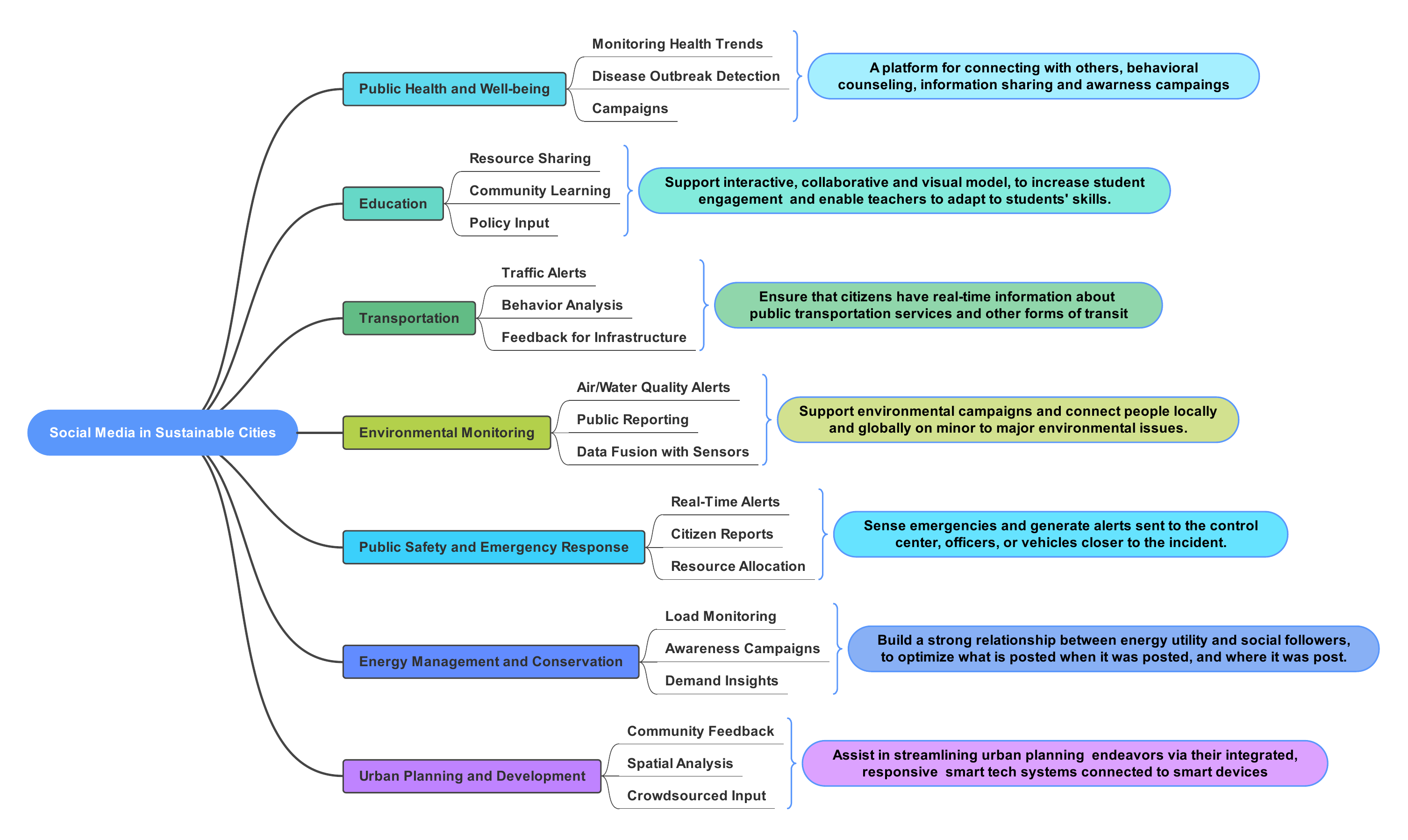}}
\caption{Applications of Social Media Analytics in Sustainable Cities.}
\label{fig:2}
\end{figure*}

In the modern world, social media has seamlessly integrated itself, significantly influencing the growth and sustainability of urban environments. Within the technology framework in developing sustainable cities, social media platforms provide a dedicated environment for people, policymakers, and organizations to connect, cooperate, and unite in marching toward shared objectives. The literature already reports the effectiveness of social media and social media informatics in resilient and sustainable cities. For instance, Yang et al. ~\cite{yang2020location} systematically examine the unique dimensions of location-centric social media data. They focus on spatial, temporal, semantic, and social aspects to address key challenges like data heterogeneity, quality, and privacy. The authors also explored the potential opportunities for urban analytics and smart city development through data fusion and analytics. This section discusses some key applications of social media informatics in sustainable cities as summarized in Figure \ref{fig:2}.

\subsubsection{Public Health and Well-being}
Real-time insights from social media informatics are revolutionizing urban healthcare. It facilitates public health officials in different tasks, such as health campaigns, controlling misinformation, keeping track of health trends, and being vigilant about disease outbreaks. For instance, social media sites like Facebook and Twitter were crucial in tracking public mood and disseminating information during the COVID-19 pandemic. 
A study by Chen et al.~\cite{chen2022multiplicity} highlights the role of social media informatics in monitoring the spread of the virus and understanding public response to health measures. Another study by Liu et al.~\cite{liu2021new} emphasizes social media data for pandemic predictive modeling in epidemiology to address health crises preemptively. Similarly, using social media informatics can significantly enhance the impact of public health campaigns and preventive healthcare strategies. Through careful analysis of public discourse, health authorities can gain valuable insights into emerging health concerns and misconceptions~\cite{madden2023constructing}. This enables them to refine their communication strategies and effectively address the needs of the public. This can allow researchers to develop targeted social media campaigns to address specific health behaviors or promote vaccination and healthy lifestyle choices. It can also effectively increase community engagement and compliance. In addition, social media platforms allow for direct engagement with the public, offering a means for authorities to address concerns, debunk misconceptions, and deliver timely health guidance~\cite{zhang2024detecting}. Collecting and analyzing vast amounts of data from social media allows for identifying at-risk populations, leading to more targeted and fair health interventions. 

However, several challenges are associated with the effective use of social media in critical applications like healthcare. Some key challenges include the processing of noisy and unstructured data, bias, misinformation, adversarial attacks, privacy and ethical concerns, and other legal and regulatory issues. Thanks to the recent developments in social media informatics, several interesting solutions have been proposed to overcome these challenges. For instance, social media informatics has been proven very effective in misinformation detection \cite{gongane2024survey}. The literature also reports several interesting pre-processing techniques for handling unstructured and noisy data for healthcare applications \cite{albalawi2021investigating}. Similarly, several interesting approaches to adversarially robust and privacy-preserving ML-based solutions have been proposed for different tasks in healthcare \cite{qayyum2022collaborative}.  However, these technical solutions are not enough to mitigate all the challenges to social media informatics in sustainable healthcare applications rather a multifaceted approach is required by involving different stakeholders, such as researchers, public authorities, physicians, and policymakers. 

\subsubsection{Education}
Social media is also playing its part in the educational sector by contributing to sustainable education and learning in several ways. For instance, social media and online social networks/platforms facilitate learners by sharing learning materials and providing resources for collaboration. The collaboration on social networks for education and learning builds global educational networks and communities, promoting lifelong learning that leads to a sustainable education system ~\cite{yigitcanlar2020sustainability}. The educational networks and communities could serve different purposes. For instance, teachers can share instructional content and motivate students to gain access to remote learning resources by using social media platforms. 
Similarly, public authorities and policymakers could use the data and insights from the data shared in these networks can use it for informed decisions and policy-making, resulting in focused educational programs or data-driven curricula that align with the community's changing requirements \cite{mekic2024data}. Educators and students could also use these networks for sharing learning materials, resources, and best learning and teaching practices, leading to more effective collaborative learning environments. 
According to Lacka et al., ~\cite{lacka2021can}, social media informatics could be very useful in determining students' needs and effective teaching tactics, leading to personalized learning experiences. The study reveals that while virtual learning can aid in achieving the goals of educational institutions with additional inputs, students relying on social media are more efficient with the strategic use of modern technologies. In addition, AI-powered social media platforms can enable peer-to-peer learning, allowing students and community members to share knowledge, provide assistance, and work together on projects outside the confines of a traditional classroom~\cite{ala2023leveraging}. Ajibade et al.~\cite{ajibade2023technological} have explored how integrating social media with e-learning platforms has improved student engagement and yielded insightful information on learning behaviors. The authors used a technology acceptance model to examine how Nigerian college students and teachers adopt social media for e-learning.

However, like other applications, social media informatics faces several challenges in the education sector. For instance, extracting relevant content is very challenging, requiring accurate social media informatics algorithms and high computational resources. Similarly, due to the lack of proper control, false information on social media could mislead learners and make access to accurate and reliable content more challenging \cite{pasieka2021harmful}. The presence of echo chambers on social media could also harm the critical thinking of students by only exposing them to content relevant to their existing beliefs \cite{lomonaco2022yes}. In addition, the digital divide, privacy and security concerns, and legal and regulatory issues also reduce the effectiveness of social media informatics in education. 

\subsubsection{Transportation}
Social media informatics also contribute to sustainable transportation in several ways. For instance, social media informatics insights could be beneficial during emergencies, such as floods, or unexpected road closures, where quick responses and updates on the road status are crucial~\cite{rinchi2024role}. The literature reports several interesting works in this direction. For instance, Ahmad et al. \cite{ahmad2019automatic} proposed an image processing framework that automatically collects and analyzes social media imagery for real-time updates on road conditions during floods. The authors mainly carried out two tasks. Firstly, images, providing evidence for road passability, are collected through a classification algorithm. The images are then analyzed for passable and non-passable roads. These applications are expected to enhance the efficiency and dependability of urban transport systems significantly. 
Moreover, social media informatics has the potential to revolutionize transportation services by studying commuter behaviors and preferences. This can result in personalized travel recommendations and improved user experiences~\cite{sun2021transportation}. Commuters' real-time data regarding traffic, public transportation efficiency, and road incidents must be analyzed to optimize urban transportation systems with social media analytics. 
Social media informatics could also be useful in gathering feedback from commuters, enabling transportation authorities to improve and adjust services based on real-time user input~\cite{kuberkar2020factors}. This enhances transportation efficiency and promotes a more inclusive and user-centered transport infrastructure in smart cities. 
The literature also reports several interesting works in this direction. For instance, Mogaji et al.~\cite{mogaji2023exploring} analyzed the effectiveness of social media for commuters with disabilities. The social media analytics made through this study greatly enhance public transportation planning and emphasize the need for innovative solutions in the unregulated transport sector.

However, several challenges, including both technical and non-technical challenges, are associated with the extraction of useful insights for sustainable transportation. For instance, it is possible that relevant data (social media posts) are not available every time, making real-time feedback on road conditions very challenging. This issue can be overcome by integrating social media data with other sources, complementing each other in such tasks. Similarly, the availability of geo-location information is crucial for effective localized insights. However, the availability of correct geo-location is not always guaranteed, making geo-localized insights more challenging \cite{mehmood2024named,alam2023role}. This challenge could be overcome by extracting location information from the social media content. For instance, Suleman et al. \cite{suleman2023floods} proposed a transformer-based Named Entity Recognition (NER) framework to extract location information/addresses from social media text. Apart from this, false information, data bias, privacy and security concerns, and other regulatory and legal issues may also reduce the effectiveness of social media informatics in transportation. These challenges require a multifaceted approach involving all the stakeholders.

\subsubsection{Environmental Monitoring and Awareness (Water and Air Quality)}
Social media also contributes to sustainable cities by improving environmental monitoring mechanisms in sustainable smart cities. People frequently report environmental issues on social media platforms, such as air pollution and water quality in their locality. This could be used by the concerned public authorities and other organizations as real-time feedback. One of the key advantages of social media informatics is its ability to analyze large volumes of unstructured social media data, including posts and comments about different environmental issues. By doing so, it can detect emerging ecological risks and determine which areas need urgent attention. These reports can be examined to find patterns and regions that require attention~\cite{salman2023review}. Using technology to ensure sustainability, resource efficiency, and improved public health helps solve critical urban issues like water, waste, air, and traffic. A recent study in~\cite{ashayeri2024exploring} used NLP to analyze geo-tagged Twitter data, finding increased negative feedback on indoor air quality in 2020 compared to 2019, emphasizing the importance of data-driven approaches for future indoor air quality management strategies. Combining social media data with conventional environmental sensors to create a more efficient monitoring system is covered in~\cite{gkontzis2024temporal}. This study analyzes time series data from Patras, Greece, identifying seasonal patterns in reported urban issues, with peaks in summer, and forecasts a consistent rise in urban problems, offering insights for informed decision-making.

Through the integration of social media informatics and traditional environmental sensors, cities have the potential to develop a monitoring system that is both dynamic and responsive. To address climate change, overpopulation, and resource misuse, the study in ~\cite{ortega2020artificial} evaluates how a medium-sized city like Granada can transition into a smart city by analyzing AI successful models, highlighting the need for strategic technological actions in areas like governance, mobility, and sustainability. Those models cannot only detect issues in real time but can also predict future environmental challenges by analyzing historical data patterns. AI can analyze geo-tagged social media posts and uncover correlations between weather conditions and pollution levels~\cite{wu2020tracking}. This valuable information can then be used to take proactive measures to mitigate any adverse effects. In addition, social media platforms can potentially increase public awareness by customizing environmental information for different communities~\cite{javaid2022understanding}. It ensures that residents receive timely updates and practical advice regarding water and air quality concerns in their areas. Such approaches enhance resource efficiency and public health, encouraging active citizen participation in environmental preservation and promoting resilience and sustainability in urban areas.

However, like other social media informatics applications, several factors make environmental monitoring and awareness in social media very challenging. Some of the key challenges include false reporting and misinformation, echo chambers, filtering large volumes of data, and data privacy and security concerns. More specifically, social media platforms and users generally focus on trendy topics, which are generally very short-lived. However, a sustainable environment requires a long-term campaign engaging social media users. 

\subsubsection{Public Safety and Emergency Response}
Public safety and emergency response is one of the key aspects of sustainable cities. Like other applications, social media informatics can play a significant role in public safety and emergency response \cite{alam2023role}. Social media allows the public and government to communicate in both directions, which is very important in emergencies. Responders may more precisely assess situations and direct resources where they are most needed by using citizen reports, location sharing, and firsthand information~\cite{imran2020using}. Authorities can promptly identify and respond to incidents, such as natural disasters, crimes, and other emergencies based on the real-time data supplied by residents. The literature indicates that social media sites such as Facebook and Twitter are critical in sharing disaster alerts, organizing rescue efforts, and giving current information to the public during emergencies. For instance, real-time tweeting during natural disasters has been used to map damaged areas and effectively allocate resources~\cite{havas2021portability}. Furthermore, social media informatics can uncover trends and anticipate probable crime hotspots, enabling preventative actions and heightened awareness in susceptible locations~\cite{butt2020spatio}. In addition to improving emergency response times, the capacity to quickly provide instructions and crowdsource information promotes community resilience and public safety system trust. In this regard, social media informatics can analyze citizen reports, geographical data, and firsthand information for practical insights, allowing for more accurate evaluation of situations and quicker response times~\cite{rafner2022mapping}. For instance, social media informatics frameworks can examine live tweets in real-time during natural calamities to dynamically chart impacted regions and propose the most effective distribution of emergency supplies~\cite{aboualola2023edge}. Moreover, it can identify patterns in social media data that could signify increasing tensions or the probability of criminal behavior. This enables law enforcement to proactively take measures and strengthen security in susceptible regions~\cite{oatley2022themes}. Guaranteeing timely and well-coordinated information and activities with the help of social media informatics enhances the efficiency of emergency responses and reinforces community resilience and public trust.

However, the extraction of meaningful insights from social media posts is very challenging. For instance, social media posts must include geo-location information to obtain localized insights. Similarly, the posts must be detailed enough to extract meaningful insights into public safety and to respond accordingly. 

\subsubsection{Energy Management and Conservation}
Energy management and conservation is one of the key aspects of sustainable cities. The literature reports several interesting solutions for efficient management and conservation of energy. For instance, Bertoldi et al. ~\cite{bertoldi2022policies} review the need for energy conservation and sufficiency policies, highlighting tools such as personal carbon allowances and progressive standards, and emphasize the importance of addressing both individual and residential energy use to complement traditional efficiency measures and achieve rapid demand reduction. Similarly, Li et al. ~\cite{li2022effective} addressed challenges such as deteriorating infrastructure, congestion, and air pollution by proposing the Social Collaboration of Renewable Energy Resources (SC-RER) technique, which maximizes local renewable energy use and the Smart Block Innovation (SBI) model for efficient community planning, achieving a 95.4\% efficiency reduction in energy consumption. Energy management and conservation initiatives in metropolitan regions can also greatly benefit from social media informatics~\cite{corbett2022tweets}. For example, social media data can be used to determine peak consumption periods for improved load control or to assess how well public awareness efforts about energy conservation are working. Through data mining for multidimensional publicity analytics, Jian et al. ~\cite{jiang2021data} demonstrate the best times to engage users and emphasize how successful social media campaigns can improve waste management results by increasing user engagement and waste collection rates. Public awareness campaigns have the potential to be significantly enhanced by leveraging social media informatics to tailor them to specific demographics, which can improve the overall impact of energy conservation efforts significantly ~\cite{fan2024role}. Furthermore, sentiment analysis, which is an application of social media informatics, can offer more profound insights into public sentiments regarding energy usage and conservation~\cite{taghikhah2022artificial}. Its potential through case studies was well documented in promoting pro-environmental behavior and managing wildfires and renewable energy, highlighting how AI-driven strategies can advance global sustainability and combat climate change.
Such analytics enable policymakers to adapt strategies accordingly to observe how social media informatics can analyze multilingual social media content and use advanced geo-location tagging to gain insights into regional energy behavior and sentiment variations.

\subsubsection{Urban Planning and Development}
Urban planning and development is the most critical aspect of sustainable cities. Unlike other applications, the role of social media informatics in urban planning is slightly different. The literature reveals that broader adoption of social media informatics in urban planning and development requires effective use of social media platforms and proper integration of crowd-sourced data obtained from social media platforms with other sources to address urbanization challenges effectively~\cite{son2023algorithmic}. These capabilities help urban planners and policymakers create more efficient and equitable cities. By incorporating public feedback into planning procedures, social media informatics can greatly assist urban planning and development~\cite{abkenar2021big}. 
Social media platforms facilitate the expression of residents' ideas regarding urban development projects, offering planners a range of perspectives and augmenting community engagement. The literature reports several cases where public feedback is used in smart city projects through social media informatics. For instance, Steinmetz et al. ~\cite{steinmetz2021liking} used data-driven analysis to examine citizen engagement with Georges River Council's social media platforms for content analysis as a part of the smart cities project.

Urban planning has the potential to go beyond conventional approaches by leveraging sophisticated algorithms to classify and give priority to community input, which can help guarantee that the perspectives of residents are not only acknowledged but also taken into account during the planning process~\cite{casali2022machine}. Furthermore, social media informatics has the potential to enhance real-time analysis and adaptive planning, allowing cities to address evolving conditions and emerging challenges swiftly~\cite{alahi2023integration}. Social media informatics has the potential to address challenges like expensive costs and limited technical knowledge, thus facilitating the implementation of participatory planning processes even in resource-limited settings. Research has demonstrated how social media data can determine public preferences and concerns about zoning, infrastructure development, and land use ~\cite{alvarado2023use}. However, practical implementation is often overlooked due to high costs, lack of political support, and insufficient technical expertise, as revealed through a case study in Mexico. Similarly, data quality, misinformation, manipulation of public opinion through adversarial attacks and echo chambers, lack of trust, data privacy and security concerns, and other legal and regulatory issues further complicate the use of social media informatics for urban planning and development.

Table \ref{tab:papers_cities} provides a summary of some of the recent works on sustainable cities' applications of social media informatics. 

\begin{table*}[]
\centering
\caption{Summary of some key papers on smart cities applications of social media informatics in sustainable cities. }
\label{tab:papers_cities}
\scalebox{.7}{
\begin{tabular}{|p{.5 cm}|p{3 cm}|p{15 cm}|}
\hline
\textbf{Ref.} & \textbf{Application} & \textbf{Overview of the Method}  \\ \hline
~\cite{yang2020location} & Urban Analytics & Examines location-centric social media data for urban analytics, focusing on spatial, temporal, semantic, and social aspects to address data heterogeneity, quality, and privacy issues. \\ \hline
~\cite{chen2022multiplicity} & Disease Monitoring & Highlights the role of social media in monitoring disease outbreaks, tracking health trends, and managing misinformation, particularly during the COVID-19 pandemic. \\ \hline
~\cite{liu2021new} & Pandemic Predictive Modeling & Emphasizes using social media data for pandemic predictive modeling in epidemiology, aiding in preemptive health crisis management. \\ \hline
~\cite{madden2023constructing} & Health Campaigns & Discusses how social media analytics can enhance public health campaigns by identifying emerging health concerns and refining communication strategies. \\ \hline
~\cite{yigitcanlar2020sustainability} & Lifelong Learning & Explores how AI in social media analytics promotes lifelong learning and community-based education, enabling educational institutions to detect emerging skills and knowledge areas. \\ \hline
~\cite{lacka2021can} & Personalized Learning & Investigates how social media analytics can personalize learning experiences by identifying student needs and effective teaching tactics. \\ \hline
~\cite{ajibade2023technological} & E-learning Engagement & Explores integrating social media with e-learning platforms to improve student engagement and gather insights on learning behaviors. \\ \hline
~\cite{rinchi2024role} & Transport Efficiency & Discusses AI-driven predictive capabilities in urban transport systems, enabling quick emergency response and optimizing transportation efficiency. \\ \hline
~\cite{zhao2022reuse} & Traffic Prediction & Highlights the role of automated platforms in providing rapid insights and predictions to address transportation challenges through social media platforms. \\ \hline
~\cite{mogaji2023exploring} & Disabled Commuter Services & Examines how social media analytics can enhance public transportation planning by offering insights into commuter preferences and interactions, particularly for commuters with disabilities. \\ \hline
~\cite{salman2023review} & Pollution Monitoring & Analyze social media data to identify patterns and regions that require attention in pollution or water quality issues, aiding in environmental monitoring. \\ \hline
~\cite{ashayeri2024exploring} & Air Quality Management & Uses NLP to analyze geo-tagged Twitter data, emphasizing the importance of data-driven indoor air quality management approaches. \\ \hline
~\cite{ortega2020artificial} & Smart City Transitions & Evaluates how AI models can be applied to smart city transitions, highlighting the need for strategic actions in governance, mobility, and sustainability. \\ \hline
~\cite{alattar2021using} & Resource Management & Explores how AI can identify public dissatisfaction with resource distribution through social media sentiment analysis, guiding proactive policymaking. \\ \hline
~\cite{abeysekera2024sentiment} & Economic Development & Analyzes social media data to identify emerging economic trends, monitor business sentiment, and support local entrepreneurs, promoting economic growth in sustainable cities. \\ \hline
~\cite{fan2024role} & Energy Conservation & Highlights how AI-powered social media analytics can enhance public awareness campaigns and improve energy conservation efforts by tailoring messages to specific demographics. \\ \hline
~\cite{son2023algorithmic} & Urban Planning & Explores the integration of AI in urban planning, emphasizing the need for collaboration and effective use of big data to address urbanization challenges. \\ \hline
~\cite{casali2022machine} & Community Feedback & Discusses how AI can enhance community engagement in urban planning by categorizing and prioritizing public feedback from social media data. \\ \hline
~\cite{hui2023greening} & Green Spaces & Investigates integrating green spaces with smart city technologies, highlighting the impact on urban livability and economic growth through sustainable practices. \\ \hline
\end{tabular}}
\end{table*}

\subsection{Social Media in Sustainable Societies}
\begin{figure*}[!t]
\centering
\centerline{\includegraphics[width=1\textwidth]{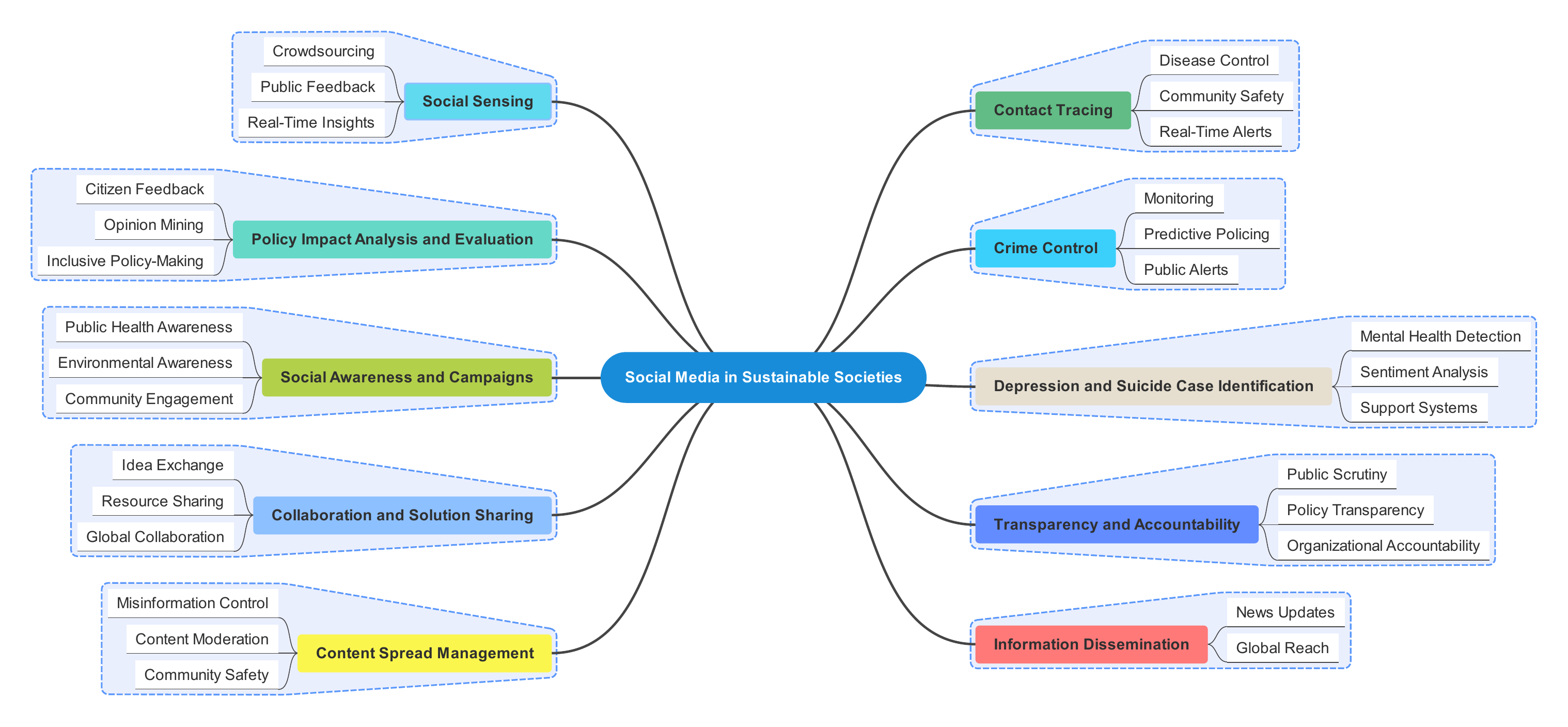}}
\caption{Applications of Social Media Analytics in Sustainable Societies}
\label{fig:3}
\end{figure*}

In the modern world, social media significantly impacts our society by directly or indirectly influencing our culture, sports, politics, and personal and collective lives \cite{kross2021social}. The literature has identified several disadvantages and negative impacts of social media on our lives \cite{appel2020social}. However, it also allows us to improve our lives by leveraging its widespread reach and interactive nature in different applications \cite{bodhi2022modelling}. This could be used to track, analyze, and disseminate useful information and practical strategies, promoting sustainability. Thanks to the recent developments in AI, NLP, and multimedia analytics, several interesting applications of social media informatics have been introduced, providing a multifaceted approach to sustainable societies. These applications collectively foster environmental, political, economic, and societal well-being. In this section, we discuss some of the social media-enabled sustainability applications. Figure \ref{fig:3} provides a summary of the social media applications in sustainable societies.


\subsubsection{Information Dissemination}
Over the last several years, social media has emerged as a faster, more accessible, and more interactive source of information dissemination \cite{westerman2014social}. Its widespread reach, worldwide accessibility, and instant access and updates are the key characteristics that make it a preferred choice for communication and information dissemination in different application domains \cite{ahmad2017jord}. The widespread use of popular social media outlets, such as Facebook, YouTube, and Twitter, expands the audience range by reaching beyond geographical bounds and making information accessible. More importantly, unlike conventional media, social media platforms have democratized information dissemination by allowing citizens (i.e., social media users) to share information. However, several challenges are associated with extracting useful information from social media. For instance, social media users generate a large volume of data containing noisy and irrelevant information on different platforms, making it very challenging to filter and analyze it for helpful information \cite{alsmadi2022adversarial}. Similarly, the content is generated in various formats, including text, images, and videos as well as in different languages as social media users tend to post in their local languages. Moreover, social media text is subject to intentional and unintentional variations, resulting in out-of-vocabulary (OOV) words. These OOV words make text processing and analysis more challenging for certain applications, such as hate speech detection \cite{huertas2023countering}. More importantly, misinformation and propaganda are the biggest challenges faced by social media as a source of information dissemination \cite{shahbazi2024social}. 

However, to cope with the challenges associated with social media data, several interesting solutions have been proposed to ensure the quality of social media data. The literature reports several interesting texts, images, videos, and multi-modal classification frameworks for automatically filtering out irrelevant content \cite{ahmad2019social}. For instance, Diaz et al. \cite{diaz2022noface} proposed an NLP framework filtering Twitter data based on the credibility of Twitter users by processing their biographies. The solution is built on the hypothesis that expert and credible social media users share relevant content to an event/topic. Similarly several solutions have been proposed to deal with OOV words generated due to intentional and unintentional textual variations in social media text \cite{khan2021enhancement, chai2023comparison}. For instance, Álvaro et al. \cite{huertas2023countering} analyzed the impact of OOV words in different applications by generating OOV words in several languages, such as English, French, German, Italian, and Spanish.  The authors then explored different ways of developing text classification frameworks more robust to intentional and unintentional textual variations. Similarly, several interesting solutions have been proposed for fake news detection and disinformation. This is one of the most active areas of research on social media informatics. For instance, Hamid et al. \cite{hamid2020fake} analyzed the effectiveness of a Graph Neural Networks (GNNs) and NLP techniques-based framework in fake news detection. All these solutions have enhanced the effectiveness of social media as a source of information dissemination. 

\subsubsection{Social Sensing: Monitoring and Reporting}
Social media, being a source of real-time monitoring, reporting, and feedback on different aspects of urban life, is also called a social sensor \cite{heglund2021social}. The basic idea behind the concept is leveraging user-generated content for meaningful insights and informed decisions in various social phenomena. The ability to generate and access content in real-time and its widespread use make it a preferred choice for crowd-sourcing to collect information from citizens on potential emerging societal and environmental issues \cite{auyb2024social}. The literature reports several situations where social media has been utilized to detect, monitor, and analyze public opinions, sentiments, behaviors, and feedback on different aspects of modern life. For instance, Ahmad et al. \cite{ahmad2022social} explored the potential of social media as a crowd-sourcing tool and source of feedback on water quality. The authors also highlighted the limitations of conventional crowd-sourcing methods, such as online and paper-based surveys. Similarly, in \cite{ahmad2019automatic} the authors analyzed social media data for real-time information on road conditions during and after floods.

However, despite the great potential, several challenges are associated with opinion mining and crowd-sourcing with social media for critical applications. For instance, the analysis of large volumes of data, data quality, and other platform-specific issues, such as the limitations and complexities of the APIs. The biggest challenge in this application is the possibility of bias in the data. This challenge could be overcome by automatically analyzing social media data/information for potential biases through social media informatics using AI and NLP techniques. The literature already reports several interesting solutions for the detection of potential biases in social media data. For instance, Raza et al. \cite{raza2024dbias} proposed an NLP framework, namely Dbias, enabling it to detect biased social media posts and identify and replace potentially biased words. Similarly, Mostatter et al. \cite{morstatter2017discovering} employed several computational and statistical methods to assess social media posts for potential bias introduced due to the social media platforms' APIs (i.e., the way the information is collected). The authors also proposed several data collection strategies to minimize bias. Social media has big potential for understanding and responding to societal and environmental issues and could be a very effective tool for crowd-sourcing if the associated challenges are tackled properly.  

\subsubsection{Policy Impact Analysis and Evaluation}
Social media could also be used as a tool for feedback, evaluation, and impact of the potential outcomes of policies by collecting real-time data from citizens. The citizens' feedback significantly impacts the policy-making process \cite{simonofski2021supporting}. One of the key advantages of social media for feedback and evaluation of policies is the diversity in opinions from different age groups, socioeconomic status, and cultural and educational backgrounds.  This diversity in social media users enhances the quality of the feedback, providing a more comprehensive and detailed view of the impact of the potential outcomes of policies. This feedback and opinions of people with a diverse background and expertise could lead to inclusive policies and decision-making processes \cite{ledesma2021popular}, which is one of the objectives of sustainable societies. This inclusive policy-making process could lead to innovative solutions to different economic, environmental, and societal issues associated with the policies. Moreover, this feedback from the citizens and inclusive decision-making process builds trust in the citizens and policymakers, leading to better engagement and contributions from the community to sustainable cities and societies. The literature reports several situations where public authorities could use social media platforms for evaluation and feedback on their policies. The feedback could be obtained in different ways. For example, polls or Q\&A sessions could be arranged on different social media platforms. Similarly, citizens' opinions could also be obtained indirectly through opinion mining or sentiment analysis of citizens' social media posts.  For instance, to support the policy-making process, Simononfski et al. \cite{simonofski2021supporting} proposed a policy analytics framework leveraging feedback of social media users by analyzing social media posts using different social media informatics algorithms. The framework also supports data extraction from other online e-participation platforms specifically designed for feedback on public policy and its impact. Similarly, Driss et al. \cite{driss2019citizens} proposed a social media informatics framework for automatically extracting useful insights from Facebook posts for policy-makers to evaluate and modify their policy by considering citizens' opinions. The framework was specifically designed for Tunisian citizens to report and give feedback on Government policies on different aspects of urban life, such as road safety and security, transportation problems, and government services. 

Despite the big potential and initial success, several ethical and technical challenges are associated with the extraction of meaningful insights for policy-makers from social media. For instance, ensuring users' privacy and complying with other regulations is critical in such applications. Similarly, all the biases in the data must be removed by ensuring the data does not represent a small portion of the stakeholders (i.e., people directly or indirectly affected by the policies). Moreover, context must be considered in processing social media posts for feedback on critical policies to avoid sarcastic comments. Sarcasm, which is one of the ways used by humans to show their emotions and convey temptation by saying the opposite of what they mean, could lead to misleading insights \cite{ahire2021sarcasm}. However, these challenges could be tackled through social media informatics frameworks for bias and sarcasm detection. In short, social media offers significant opportunities to enhance and make the policy-making process more inclusive by considering citizens' feedback if handled properly. 


\subsubsection{Social Awareness and Campaigns}
Over the last several years, social media has emerged as a powerful tool for social awareness and campaigns. Social media possesses several key characteristics, such as worldwide accessibility and widespread use, interactive nature, and ability to engage large and diverse audiences, which make it a preferred choice for raising social awareness and conducting campaigns. Moreover, the freedom of sharing multi-modal content in terms of text, images, videos, and audio clips helps social media users create engaging and appealing content for social awareness and campaigns. Similarly, it also allows the opportunity to collaborate with social media influencers, benefiting from their popularity and large followings to amplify campaigns. The literature reports several studies where the large followings of social media influencers were used to amplify the campaigns \cite{kostygina2020boosting}. Moreover, several social media platforms, such as Facebook, YouTube, and Instagram, also provide an opportunity for real-time engagement via live streaming and webinars. 

The literature reports several successful social awareness campaigns through social media including public health and safety, environmental, political, heritage preservation, educational, and animal and human rights campaigns. For instance, Al-Dmour et al. \cite{al2020influence} explored the potential of social media in public health campaigns during the COVID-19 pandemic by analyzing data from multiple social media platforms in Jordan. The authors reported a significant positive influence on public health protection against COVID-19 by building awareness and motivation for behavioral changes to cope with the pandemic. Similarly, Zamani et al. \cite{zamani2024use} utilized social media platforms and Augmented Reality (AR) technology for road safety awareness. To this aim, a questionnaire and AR advertisements on road safety were shared on Facebook, which were shared with more than 200,351 people and obtained  719,296 impressions and 3,218 clicks. More recently, Laban et al. \cite{laban2024role} analyzed the role of TikTok in the dissemination of the Palestinian narrative during the war on Gaza. Chauhan et al. \cite{chauhan2016social} examined the success of social media in LGBT campaigns. Similarly, Hajri et al. \cite{hajri2024role} and Liu et al. \cite{liu2015large} analyzed the effectiveness of social media in environmental and educational campaigns, respectively. 

All these studies reported a significant positive influence of social media on public awareness campaigns. However, similar to other social media applications, several challenges are associated with conducting successful social awareness campaigns on social media. One of the biggest challenges in this regard could be the so-called echo chambers on social media \cite{terren2021echo}, which could limit the reach and impact of campaigns to people outside the echo chambers. Similarly, the abundance of misinformation on social media can also undermine the credibility of these campaigns \cite{alrashdi2022impact}. However, these challenges can be overcome through social media informatics by developing effective misinformation detection and adversarial robust frameworks to break the echo chambers \cite{mahmoudi2024echo}. We have already discussed the literature on misinformation and fake news detection in the above sections. The literature also reports interesting attempts and debates on breaking the echo chambers on social media. However, this requires a joint effort from different stakeholders to develop rules and regulations to ensure the core values of democratic societies, such as freedom of expression and user privacy \cite{avin2024impossibility}.


\subsubsection{Collaboration and Solution Sharing}
Social media also provides a platform for collaboration and solution-sharing to address different societal challenges. For instance, social media platforms like Twitter, Facebook, and LinkedIn could be used to discuss and gather ideas and potential solutions to various societal issues from a diverse audience \cite{zaffar2012knowledge}. The literature reports several studies on effective communication and sharing solutions for different issues \cite{nugraha2021enterprise}. For instance, Okonkwo et al. \cite{okonkwo2023role} explored the potential of social media as a source of communication, collaboration, and sharing solutions for potential challenges in businesses. Thomas et al. \cite{thomas2013social} analyzed and reported the importance and the effective role played by social media for collaboration within an organization. Considering the effectiveness of social media in collaboration and solution sharing, Governments, public authorities, and non-government organizations also rely on social media, seeking potential solutions and collaborations on different environmental and societal issues. For instance, The Spanish government used social media to engage citizens and public authorities in combating COVID-19 \cite{criado2020using}. Similarly, the German and Estonian governments launched civic hackathons, seeking innovative potential solutions from the citizens for handling the COVID-19 crisis \cite{criado2020using}. Social media platforms were also used for seeking internal and external collaborations and expert opinions on environmental challenges in Pakistan \cite{syed2024impact}.

Social media platforms can also promote collaboration by facilitating the gathering of funds and resources for different projects, solving societal challenges \cite{bhati2020success}. These platforms possess several characteristics, facilitating fundraising in various ways \cite{mihai2018usage}. For instance, its widespread use allows fundraisers to reach global audiences. Similarly, social media platforms support multi-modal data, allowing the creation of more appealing textual and visual content \cite{hassan2019sentiment}. Similarly, several social media platforms, such as Facebook, support integrated and third-party fundraising tools, such as GoFundMe \footnote{GoFundMe: https://www.gofundme.com/}, and Patreon \footnote{Patreon: https://www.patreon.com/}, facilitating the creation, and promotion of fundraising campaigns as well as tracking the progress. Several studies have reported the success of these fundraising tools in real-life fundraising campaigns for healthcare and treatment \cite{ren2020understanding,song2020use} and disaster relief \cite{radu2019bridging}. 

However, despite the significant success, social media-based fundraising companies face several challenges. The biggest challenge to such campaigns is the trust and credibility issues. Social media is prone to scams and fraud that could make potential donors wary of contributing \cite{zenone2019fraud}. But, thanks to social media informatics, the detection of fraudulent fundraising campaigns is possible by deploying ML and NLP algorithms. The literature already reports the feasibility of such ML and NLP-based solutions for the detection of fraudulent fundraising campaigns. For instance, Perez et al. \cite{perez2022call} proposed a multi-model framework for the detection of fraudulent fundraising campaigns using an ML model trained on textual and visual features extracted from fundraising campaigns. To summarize, social media has a significant potential for tackling societal issues by facilitating collaboration and sharing solutions among different stakeholders.

\subsubsection{Transparency and Accountability}
Social media possess several characteristics that make it a suitable tool for promoting transparency and accountability in different aspects of sustainable cities and societies \cite{treem2015social}. For instance, real-time communication and instant access to information allow all stakeholders including public authorities, businesses, and non-profit organizations to share information, and provide updates on their policies, decision-making processes, and actions, promoting transparency. Similarly, the general public can give feedback and share their opinion on the services, facilities, policies, and their potential impact on the public and environment. The interactive nature of the communication provided by social media also allows organizations to respond to the citizens' feedback and highlight the changes made in response to the feedback \cite{einwiller2015handling}. Similarly, social media platforms facilitate crowd-sourced investigations and encourage citizens to investigate and verify information provided by the concerned authorities, businesses, and organizations \cite{stamati2015social}. The literature also reports on the effectiveness of social media as a medium for reporting potential corruption, fraud, or abuse of power cases \cite{srivastava2016role}. 

However, similar to the other applications, several challenges are associated with promoting transparency and accountability through social media. Firstly, social media is prone to misinformation and fake news, and the data shared on its different platforms must be verified \cite{olan2024fake}. Secondly, there is no accountability for social media users, and deliberate and coordinated campaigns could be launched to defame organizations or individuals by creating illusions and manipulating public opinions \cite{akhtar2024sok}. Moreover, considering the abundance of misinformation and fake news, a vast majority of citizens do not trust social media content, which could further hamper transparency and accountability efforts. 

These challenges could be overcome with social media informatics algorithms. As discussed earlier, the literature reports several interesting solutions for disinformation and fake news propaganda detection in social media platforms. Similarly, the authenticity of shared documents could be proved through multimedia forensics. Multimedia forensics has been proven very effective in assessing the credibility of multimedia information shared in social media platforms \cite{pasquini2021media}. Social media has great potential for promoting transparency and accountability, however, a multifaceted approach involving different stakeholders to combine technological solutions, regulatory/legal aspects, and community engagement strategies, is required to overcome the associated challenges.

\subsubsection{Depression and Potential Suicide cases identification}
Social media users, generally, tend to share their emotions by posting on social media, and these posts provide explicit and implicit cues including textual, visual, and contextual cues about their emotions and mental status. In terms of text, several words, such as ''happy'', ''excited'', ''angry'', ''frustrated'', and ''disappointed'', could hint at the mental state of social media users. Social media users also use text formatting (e.g., upper-case words) or punctuation to express the intensity of their emotions \cite{prikhodko2020ways}. In visual content, cues for emotion detection include facial expressions, color combination, background color, different objects in the images/frame, body language, and situational context \cite{wei2024learning}. Several social media platforms support emojis, which social media users use to show their emotions \cite{cherbonnier2024people}. These platforms provide emojis for positive and negative emotions, reflecting their sentiments, moods, and mental state.   

Social media informatics allows the extraction of these cues of emotions and sentiments and predicts social media users' sentiments and mental states. The literature reports several interesting social media informatics algorithms and frameworks for predicting users' mental status from their posts \cite{saxena2020emotion}. These frameworks are useful in the early identification of potential depression and suicidal cases, allowing for timely intervention and support \cite{lin2020sensemood}. However, this requires continuous monitoring of individuals by analyzing the timing and frequency of their posts with explicit and implicit cues of emotions, and alerts to be generated when there is a sudden change in posting frequency during stressful times (e.g., a sudden increase in posts with negative emotions or lengthy inactivity). 

However, several technical and ethical challenges are associated with this continuous monitoring and data processing for depression detection. For instance, a fine balance between privacy and the need for data analysis needs to be defined to avoid the violation of the basic human right of privacy \cite{garcia2021socially}. Moreover, it requires quality data and accurate interpretation of the context of the posts. The context of the posts is very critical in this application as the context in which certain textual, visual, and behavioral cues are used can significantly improve the accuracy and reliability of depression and potential suicidal cases.

Thanks to the recent development in social media informatics, contextual information could be automatically extracted from the threads of social media posts. The literature reports several interesting frameworks for the identification of depression and potential suicidal cases in social media posts by considering the emotion cues and the context in which they occur \cite{william2021text}. For instance, Tejaswini et al. \cite{tejaswini2024depression} proposed a hybrid deep learning neural network-based framework, composed of a Convolution Neural Network (CNN) and a Long Short-Term Memory (LSTM) model, for combining the contextual information along the textual cues. The authors also employed several NLP techniques to clean and simplify the social media posts before processing them by dealing with OOV words. Thekkekara et al.  \cite{thekkekara2024attention} also employed a hybrid model with an attention mechanism on CNN-BiLSTM extracting contextual and linguistic features from social media posts to identify potential depression cases. To summarize, social media informatics has been proven very effective in the early detection of depression and potential suicide cases. However, human intervention and interpretation are crucial in this application, especially for contextual information and understanding of sarcasm, which are still very challenging tasks for social media informatics algorithms.

\subsubsection{Crime Control}
Crime control is another sustainable society application where social media plays its part. Social media informatics could be useful in crime control in several ways. On one side, it could be used by public authorities as a source of real-time monitoring, communication, dissemination of information about crimes, and generating alerts and warnings on a larger scale \cite{peters2019social}. Similarly, with its ability to engage the audience in crowd-sourcing, the public can share useful information with the concerned authorities by sharing real-time updates, photos, and videos of the crime scene \cite{domdouzis2016social}. Moreover, thanks to the recent development in social media informatics, social media could also be useful in the identification of potential suspects through behavioral analysis and social media profiling. This involves extracting critical information from social media profiles including their posts, activities, connections, locations, and behaviors \cite{waszkiewicz2024s}. Social media profiling could also be used in predictive
policing, allowing the identification of crime hot spots using data analytics and AI algorithms. This predictive policing is very helpful in preventing potential crimes by allocating more resources to crime hot spots \cite{singh2024advancing}. As one of the hot topics of research in social media informatics, several aspects of predictive policing have been explored in the literature including both the positive and negative  \cite{mcdaniel2021predictive}. For instance, Abbas et al. \cite{abbas2024predicting} examined and reported the effectiveness of social media profiling and predictive policing in investigation and crime control in the Sargodha district of Pakistan. Similarly, Vella et al. \cite{vella2024criminal} also examined the effectiveness of predictive policing and social media profiling for crime control in Malta. The authors reported the effectiveness of the tool in conjunction with other investigation methods for better investigation and crime control. Several studies have also highlighted the negative aspects of social media profiling and predictive policing by providing detailed risk assessments \cite{berk2021artificial}. For instance, Alikhademi et al. \cite{alikhademi2022review} examined the potential of predictive policing from the fairness point-of-view. The literature reports several cases where predictive policing algorithms have been found biased against certain races and genders \cite{ahmad2022developing}.

Some of the key challenges associated with the use of social media informatics for crime control and predictive policing include privacy concerns, fairness and accountability, and bias. Social media informatics could be proved very effective in crime control if these challenges are handled properly.  

\subsubsection{Contact Tracing}
Contact tracing, a tool/mechanism for controlling the spread of infectious diseases, is another critical application of social media informatics. It involves the identification of citizens who have come in contact with infected individuals with communicable/infectious diseases, such as COVID-19.  Contact tracing has been proven very effective in controlling the spread of several diseases, such as COVID-19, Tuberculosis (TB), and Ebola, where the individuals were advised to quarantine or other necessary precautions were taken, significantly reducing the chance of further transmission \cite{hossain2022effectiveness}. The commonly used methods for contact tracing include manual contact tracing, digital contact tracing, and community engagement. In manual contact tracing, interviews of infected individuals are usually carried out, enquiring about their potential contact during a specific period. The contacts are then notified and advised accordingly. However, manual contact tracing at a larger scale is a time-consuming and tedious job. The limitations of manual contact tracing could be overcome through digital contact tracing \cite{trivedi2020digital}, where mobile applications are developed to trace contacts of infected individuals using Bluetooth technology. The mobile applications also possess an alert feature, automatically notifying the individuals who came in close contact with the infected individuals. These applications have been proven very effective in contact tracing at a larger scale \cite{trivedi2020digital}. Several works reported the effectiveness of these digital contact tracing applications during the recent COVID-19 pandemic \cite{ahmad2021sentiment}. However, these applications also have several limitations. Ahmad et al. \cite{ahmad2022global} analyzed the performance of 46 COVID-19 contact tracing applications used in different parts of the World by sentimentally analyzing users' feedback on Google Play and Apple’s App Store. The key issues highlighted by the users include privacy concerns, battery consumption, difficulties in installation and updates, and frequent crashes.    

Social media could also be used as an alternative for contact tracing. Social media brings several advantages to contact tracing. For example, it could lead to a more speedy identification and notification of contacts compared to manual contact tracing. Similarly, this could engage large audiences compared to digital tracing without developing additional applications. More importantly, this will engage citizens in contact tracing through crowd-sourcing. The literature already reports success stories of social media in contact tracing. For instance, Wang et al. \cite{wang2020new} proposed and evaluated the performance of a digital contact tracing application by integrating GPS and social media. Similarly, Sahraoui et al. \cite{sahraoui2022traceme} also developed an online social networks-based contact tracing application for tracing COVID-19-infected individuals. However, similar to other digital contact tracing applications, social media-based contact tracing faces several ethical and technical challenges in the form of privacy concerns, informed consent, and misinformation.

\subsubsection{Content Spread Management}
Social media content spread management refers to mechanisms utilized to manage and mitigate the spread of harmful, false, or provoking content and behaviors within society \cite{vosoughi2018spread}. In building a sustainable society, mechanisms for spread control must be built so that misinformation, harmful content, or any other form of negativity cannot stand in the way of social coherence, environmental protection, and economic stability \cite{wardle2017information}. Content spread control is essential to a safe and trusted online social media environment \cite{muhlmeyer2021information}. Effective social media content control reduces the risks related to harmful, misleading, inappropriate, or personal content, ensuring the safety and privacy of users and maintaining trusted online social media environments \cite{gillespie2018custodians}. The major components of content spread management include malicious content detection/classification, users' behavior estimation, content impact assessment, and content control mechanisms \cite{chandrasekharan2017bag}. These components significantly contribute toward ensuring that online platforms remain safe, trustworthy, and conducive to positive interactions.

Malicious content detection involves identifying and flagging content that could negatively influence individuals or communities, such as hate speech, offensive, racism, misinformation, cyberbullying, violent content, and other material \cite{schmidt2017survey}. Identifying malicious content involves various methods, from manual identification to automatic methods. The manual identification of malicious contents can be achieved by expert review \cite{gerrard2020behind} or community reporting \cite{chandrasekharan2017you}. Although expert review-based content identification can be highly accurate, the existing social media platforms and research community rely on community-based content labeling due to the high cost and time required to label such a high volume of social media data. Community Notes, a feature launched by Twitter (now X), leverages community reporting to identify, assess, and offer context or corrections to misleading or harmful content. Studies reported the effectiveness of this tool in combating misinformation about the COVID-19 vaccine \cite{allen2024characteristics}. Community Notes are perceived as more trustworthy than traditional information tagging mechanisms by users; however, this tool is also susceptible to malicious manipulation by a group of people. For instance, a post by a follower of one political party can be flagged as misinformation by other political groups.  

Besides manual methods, numerous studies have reported the effectiveness of automatic malicious content detection using AI techniques. These automated systems analyze posts in real-time, identifying harmful or inappropriate content based on predefined criteria or learned patterns from extensive training datasets \cite{schmidt2017survey}. These tools quickly flag or remove malicious content, improving user safety by minimizing the malicious content spread. Automated content moderation complements human oversight by addressing the limitations of human moderators in handling enormous data volumes while keeping consistency and moderator bias at a low-level \cite{gillespie2018custodians}. However, it is the effective balance of automated systems and human judgment in adjusting subtle language and context so that false positives do not arise and free expression is enabled \cite{gorwa2020algorithmic}.

Content control and moderation are not simple features for social media platforms like Facebook, X, and Instagram. Mark Zuckerberg \footnote{Big Tech Needs More Regulation: https://about.fb.com/news/2020/02/big-tech-needs-more-regulation/} states that “platforms like Facebook have to make trade-offs ... between free expression and safety” and that there is rarely a clear “right” answer. Combining automated and human moderation techniques can enhance content control; however, more sophisticated and tangible content control and moderation methods are needed to ensure a trustworthy social media environment. Sarah Myers West \cite{myers2018censored} reported that content control and moderation can significantly impact users emotionally and practically. Users expressed a desire for more transparency and human interaction in moderation, often feeling that current systems are impersonal and difficult to navigate. Moreover, platforms should consider educational approaches that aid users in understanding and complying with community guidelines rather than depending solely on corrective measures.


Table \ref{tab:papers_socities} provides a summary of some of the recent works on the various applications discussed above.
\begin{table*}[]
\centering
\caption{Summary of some key papers on the applications of social media informatics in sustainable societies.}
\label{tab:papers_socities}
\scalebox{.7}{
\begin{tabular}{|p{.5 cm}|p{4 cm}|p{15 cm}|}
\hline
\textbf{Ref.} & \textbf{Application} & \textbf{Overview of the Solution}  \\ \hline
 \cite{zhu2018information}& Information Dissemination & Proposes an information dissemination model in social media by focusing on the relationship between the original and updated information, then proposing the priority and forecast trends in information dissemination. \\ \hline
 \cite{ahmad2017jord} & Social Sensing: Monitoring and Reporting & Proposes a social media informatics framework for automatically collecting and analyzing (classification into relevant and irrelevant data) disaster-related multimedia content from various social media platforms. \\ \hline
 \cite{ahmad2022social} & Social Sensing: Monitoring and Reporting &  Proposes a social media informatics framework for water quality analysis by automatically collecting and analyzing social media users' feedback on water quality using several text analysis and topic modeling techniques.\\ \hline
\cite{gal2014impact} & Policy Making, Impact Analysis, and Evaluation &  Explores the potential of social media informatics in making transport policies by analyzing how the individual transport suppliers can utilize social media data and potential contribution to the policy development based on transport-related information by the public.\\ \hline
\cite{driss2019citizens}&Policy Making, Impact Analysis, and Evaluation &  Proposes a social media informatics framework for extraction of relevant data from social media platforms to facilitate government policy-makers using semantic text analysis. \\ \hline
\cite{scholtz2016social}& Social Awareness and Campaigns &  Designed a campaign backed by a theoretical model for environmental awareness of staff members in a Higher Education Institutions of South Africa. A centralized website/application along with various social media forms (surveys and questionnaires), were used for the campaign. \\ \hline
 \cite{bonnevie2020using} & Social Awareness and Campaigns & Conducted a study to assess the impact of social media awareness campaigns on flu vaccination, and analyze shifts in social norms about flu vaccination after the campaign. Social media influencers were also involved in the campaign by allowing them to create and share their content on their social media pages. \\ \hline
\cite{leka2023spot}& Collaboration and Solution Sharing & Proposes a social media-centered responsive application with a user-friendly interface for Albanian students to collaborate on various academic-related
topics.  The application provides a responsive collaborative environment among e-learning participants where they can share their ideas and solutions for different problems.  \\ \hline
\cite{stamati2015social} & Transparency and Accountability & Conducts a study to analyze the effectiveness of social media for enhancing citizens' participation and engagement in
facilitating openness, transparency, and reducing corruption. \\ \hline
\cite{ghosal2023depression}& Depression and Potential Suicide cases identification &  Proposes an NLP framework for the identification of depression and suicidal risk content by employing multiple NLP models including fastText embedding and TF-IDF features, which are then used to train XGBoost for accurate text classification.\\ \hline
\cite{angskun2022big} & Depression and Potential Suicide Cases Identification & Proposes a social media informatics framework for real-time depression detection by analyzing both demographic
characteristics and their Twitter posts for two months after having
answered the Patient Health Questionnaire-9. Several ML algorithms including SVMs, Decision Tree, Naïve Bayes, Random Forest (RF), and DL algorithms are used for the analysis. \\ \hline
\cite{yang2018crimetelescope} & Crime prediction/control & Propose a crime hotspots detection platform, namely CrimeTelescope, relying on multi-modal data from social media platforms. The platform continuously
collects and extracts key features from social media data using statistical and linguistic algorithms for crime hotspot detection and mapping. \\ \hline
\cite{tam2023multimodal}& Crime Prediction/control & Proposes a fusion-based social media informatics framework, combining multi-modal data, for crime predictions. Individual feature vectors are extracted from each type of data using a ConvBiLSTM model, which are then fused into a single feature vector for classification/prediction. \\ \hline
\cite{sahraoui2022traceme}&  Contact Tracing & Proposes a social media-based contact tracing application for the prevention of COVID-19, where contacts are detected in real-time using
traditional proximity approaches (i.e., Mobile and Wireless Networks and social networks). The potential contacts are then notified using social networks. \\ \hline
\cite{chen2007incorporating} &  Contact Tracing & Proposes a contact tracing platform by incorporating geographical contacts into social network analysis where personal
and geographical contacts are used for the construction of contact networks and performing connectivity analysis for the identification of potential contacts of the patients. \\ \hline

\end{tabular}}
\end{table*}
\section{Challenges and Potential Solutions}
\label{sec:Challenges and Potential Solutions}
\begin{figure*}[!t]
\centering
\centerline{\includegraphics[width=0.8\textwidth]{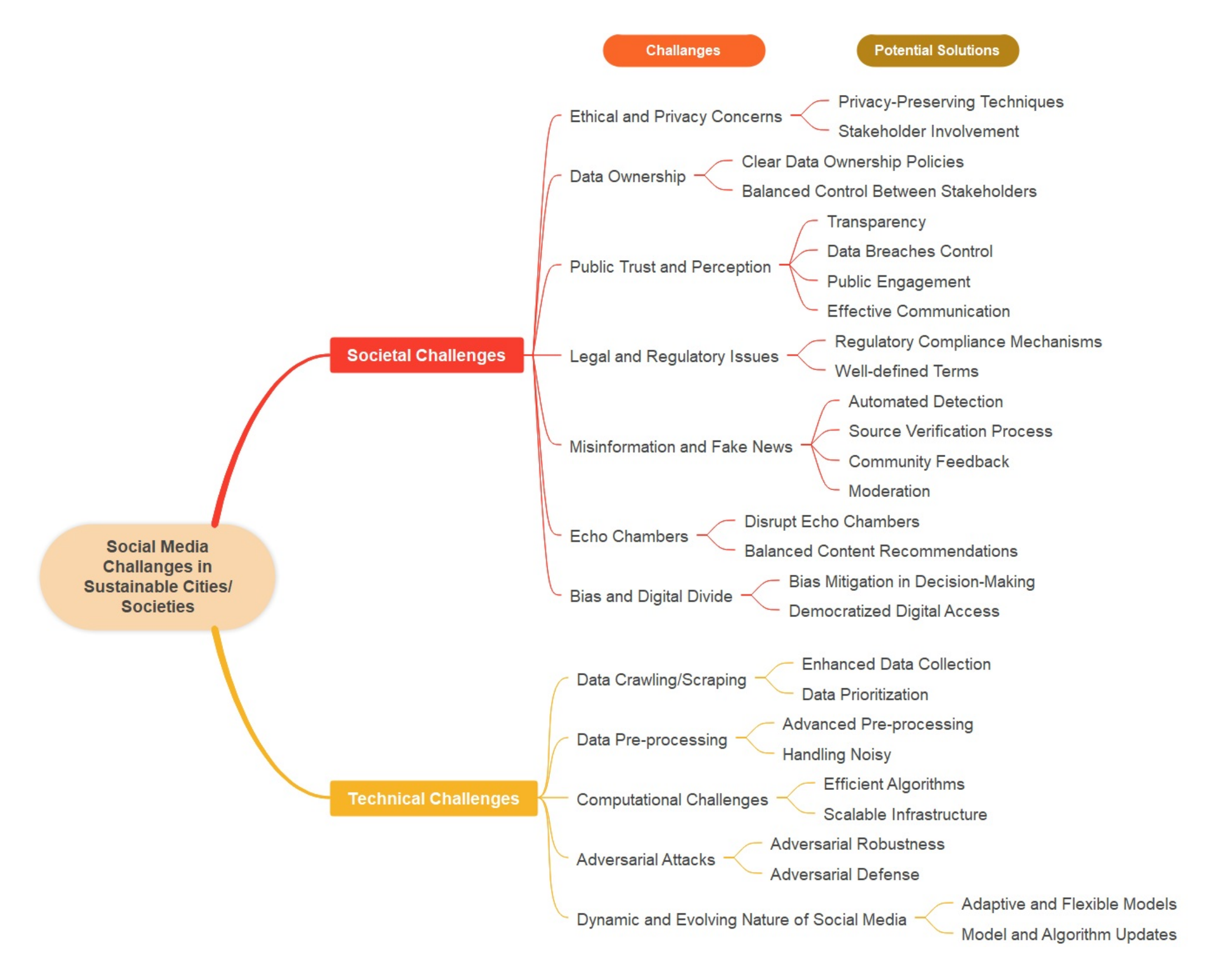}}
\caption{A summary of the challenges and potential solutions.}
\label{fig:4}
\end{figure*}

In Section \ref{sec:applications}, we discussed some of the challenges specific to different applications. This section discusses some common key challenges associated with social media informatics for sustainable cities and societies, which are summarized in Table \ref{tab:keychallenges}. Figure \ref{fig:4} provides a graphical abstract of these challenges and their potential solutions. For better organization of this section, these challenges are divided into two categories, namely (i) societal challenges and (ii) technical challenges. 

\subsection{Societal Challenges}
Several societal challenges are associated with social media informatics for sustainable cities and societies. These challenges significantly limit its effectiveness in achieving sustainable cities and societies and need a joint and multifaceted approach involving all the stakeholders to be tackled. In this section, we discuss these challenges and their potential solutions.

\subsubsection{Ethical and Privacy Concerns}
Social media informatics for various sustainable cities and societies' applications involves the risk of leakage or potential misuse of the users' information, which raises ethical and privacy concerns \cite{ahmad2022developing}.  Overcoming these concerns involves several factors and complexities, such as informed consent, anonymization, removal bias, data management \& protection, compliance with regulations, and ensuring transparency and accountability in case of data privacy breaches and information leakage.  

 Informed consent, which ensures that social media users whose data have been used must be informed about the data usage, application, and risks associated with the potential misuse, is one of the major challenges. Social media informatics applications of sustainable cities and societies generally undergo several changes, and it becomes difficult to give prior commitments about the potential future use of the data \cite{ahmad2022developing}. Similarly, new applications could emerge over time and the data may be used solely or merged with other information sources making informed consent even more challenging \cite{hand2018aspects}. Moreover, informed consent has four different conditions \footnote{Everything You Need to Know About Informed Consent: https://humansofdata.atlan.com/2018/04/informed-consent/}. Firstly, it must be ensured that the social media users whose data is used must be provided with all relevant information. Secondly, the authorities/application developers must ensure that the users completely understand the relevant information. Moreover, participation should be voluntary, and the participants must be able to decide themselves whether their data should be used by considering the risks involved with it. The fulfillment of these four conditions is very challenging. Another ethical and privacy concern about social media informatics applications for sustainable cities and societies is the risk of potential leakage of sensitive information about social media users, which may expose them to various threats. To avoid such issues, the data must be completely anonymized. However, data anonymization techniques have several limitations, and there still is a risk of re-identification of the participants \cite{pawar2018anonymization}, raising severe ethical concerns. 

Bias is also one of the most critical challenges to social media informatics that directly affects the decision-making process. Bias in any form (e.g., algorithmic and data bias, representation bias, etc.,) could compromise on the merit of the decision-making process, leading to unfairness and injustice \cite{kontokosta2021bias}. Removing bias in social media informatics requires a multifaceted approach, mainly due to various sources of bias, by ensuring inclusive and transparent design of algorithms and data collection strategies.  Moreover, transparency and a proactive approach are necessary to address the consequences of data breaches. For instance, the information about the data breach, such as the type and extent of the data breach, the affected users, the potential consequences, and the way forward, must be disclosed clearly.  

Overall, addressing the ethical and privacy concerns associated with social media informatics applications for sustainable cities and societies requires a multifaceted approach by involving different stakeholders, such as ethicists, legal experts, and social scientists, and maintaining a balance between the benefits of social media informatics and the protection of individual rights. This could be achieved through state-of-the-art solutions for data collection, protection, and anonymization, informed consent, bias detection, and mitigation, and ensuring transparency and accountability.  

\subsubsection{Data Ownership}
Data Ownership is another critical aspect of social media informatics especially in sustainable cities and societies' applications where a large amount of user-generated social media content is processed. Data ownership is a complex phenomenon and requires several ethical aspects of data collection, usage, control, and monetization of the data, to be considered \footnote{Smart City Data Ownership: https://tinyurl.com/3a3uvthb}. Generally, social media informatics applications for sustainable cities and societies involve multiple stakeholders, such as public authorities, private businesses, non-profit organizations, and the general public, with different goals and motivations. For instance, public authorities may aim to facilitate their citizens through sustainability projects. On the other hand, private businesses involved in the applications and services might be more interested in monetization of the applications/services and the associated data. This contrast in the motivations of the stakeholders makes data ownership a more complex phenomenon \cite{ahmad2022developing}. In such scenarios, defining and clarifying data ownership in sustainable cities and society applications is very critical for the success of the applications, and also helps in ensuring the privacy and rights of the citizens as the monetization of the data could compromise the privacy of the citizens \cite{ahmad2022developing}. However, businesses invest in such applications and more control of the citizens (whose data is processed) over the data would harm their interests as they will decide how the data is used, shared, and monetized, which may reduce interest in the investment from the businesses. A fine balance between the business and citizens' control over the data is critical for the success of sustainable cities and societies applications/projects.

\subsubsection{Public Trust and Perception}
Public trust and perception are very critical for the success of social media informatics applications especially in human-centered applications, such as healthcare, public safety, education, crime control, resource management, and public services. This involves people's trust, opinion, and perception of the data collection, processing, usage, and potential risks \cite{zhang2023we}. In sustainable cities and societies, public trust and perception of social media and informatics are among the biggest challenges, significantly affecting the effectiveness of social media informatics. This could affect social media informatics for sustainable cities and societies in several ways. For instance, social media users will feel reluctant to share information on social media platforms if they don't trust the way their data is handled. Similarly, the public will also not trust the information and campaigns conducted over social media, which may halt several interesting and innovative sustainable cities and societies' applications or impact their effectiveness. 

To overcome these issues, a multifaceted approach is required to build and maintain public trust in social media informatics through transparency, accountability, ethical practices, public engagement, and effective communication. However, for the potential solutions to these challenges, it is important to understand the factors that contribute to the erosion of public trust in social media informatics. In this regard, several factors, such as growing privacy concerns (mainly due to several data breach incidents and lack of control over the data), the abundance of propaganda, misinformation, and fake news, and data manipulation through data moderation, contribute to this lack of trust and perception of social media informatics and its applications \cite{hsieh2021can}. Mitigation of these factors will not only ensure individuals' rights and privacy but also build public trust in social media informatics, allowing social media informatics to contribute to the objective of sustainable cities and societies.


\subsubsection{Legal and Regulatory Issues}
Social media informatics for sustainable cities and societies also faces several legal and regulatory issues. The legal and regulatory policies though are useful in ensuring social media users' privacy and rights, they add overhead to the deployment of social media informatics for various applications. For instance, several legal and regulatory frameworks worldwide, the "General Data Protection Regulation (GDPR) \cite{voigt2017eu}" and the "California Consumer Privacy Act (CCPA) \footnote{CCPA: https://www.oag.ca.gov/privacy/ccpa}", are designed to protect users' privacy by defining rules and regulations regarding data collection, processing, and sharing, making it challenging and time-consuming to follow the rules, such as informed consent and anonymization, etc. Similarly, the regulations on data ownership, which determine the ownership and control over data used for social media informatics, also make it very challenging to use the data for new applications, especially where multiple stakeholders are involved, without informed consent \cite{ahmad2022developing}. Other legal and regulatory concerns associated with social media informatics for sustainable cities and societies include transparency and accountability, safety and surveillance, and privacy \& confidentiality \cite{bhatia2019ethical}, making processing social media data more challenging. To address these challenges, a fine balance between social media data usage and the implementation of legal and regulatory laws is required.   

\subsubsection{Misinformation and Fake News}
Social media has been one of the major sources of propaganda, misinformation, and fake news, as its various platforms are prone to misinformation and fake news \cite{suarez2021prevalence}. Several factors contribute to the abundance of misinformation and fake news \cite{chen2023spread}, such as widespread usage and worldwide availability, ease and freedom of content creation and dissemination, no central control, monetary incentives, easy manipulation of information and public opinions, and lack of proper verification mechanisms. This poses significant challenges for ensuring the accuracy and authenticity of the information shared on social media. The literature already reports several interesting solutions for misinformation and fake news detection exploring different aspects of social networks using NLP, multi-modal, and graph-based techniques \cite{aimeur2023fake}. Apart from fake news detection, several forensic methods have been proposed for the verification of news shared on social media platforms in the form of text, images, and videos \cite{pasquini2021media}. However, verification of information shared on social media is still an open challenge. To further enhance the accuracy and authenticity of information shared on social media a multifaceted approach, involving social media users in the verification process along with the AI-based solutions, is required. It is also the moral responsibility of social media users to verify and check sources of information/news before sharing it. Moreover, policies and regulations must be implemented \cite{glaeser2010regulating}, allowing strict action against misinformation spreaders as well as social media platforms failing in news/moderation and taking appropriate actions against the individuals, organizations, communities, and groups involved in propaganda and misinformation. 

\subsubsection{Echo Chambers}
Another key challenge social media informatics faces these days is the presence of echo chambers. These echo chambers refer to situations where someone is interested in information and opinions similar to their thoughts and opinions and is reluctant to consider other opinions and information \footnote{What is an echo chamber? https://edu.gcfglobal.org/en/digital-media-literacy/what-is-an-echo-chamber/1/}. Echo chambers are very common in social media, created intentionally or unintentionally \cite{zimmer2019echo}. The intentional creation of echo chambers on social media could be motivated by several factors including spreading propaganda and misinformation, reinforcing their beliefs and opinions, influencing public views for political, religious, business, and other motives, and censorship and control over social media content. These intentionally created echo chambers may involve various groups and organizations, such as political and religious groups, media and news outlets, non-government organizations, extremist groups, and other ethnic groups and communities. These echo chambers operate in various ways \cite{cinelli2020echo}. The automatic recommendation systems play their part by recommending specific content to social media users based on their preferences \cite{cinus2022effect}. This engages individuals with certain types of content, leading to frequent exposure to similar viewpoints and beliefs. Similarly, behavioral analysis algorithms contribute to the exposure of social media users to certain content based on their activities over social media \cite{moe2023polarisation}.   

Echo chambers may also have some positive aspects, and could contribute to sustainable society and city applications. For instance, these echo chambers could help in building communities and groups for certain societal challenges, such as climate change. However, dismantling these echo chambers is crucial to ensure fairness, transparency, and justice in society \cite{schoeffer2023online}. To cope with this critical challenge to social media informatics, a multi-faceted approach is required by engaging all stakeholders and deploying state-of-the-art techniques and changes to users' behavior.  

\subsubsection{Digital Divide}
The digital divide in society (i.e., the gap among individuals, groups, and communities in access to social media and other relevant technologies) also poses several challenges to social media informatics for sustainable cities and societies \cite{dargin2021vulnerable}. One of the biggest challenges in this regard is the potential bias in decision-making due to AI algorithms trained on skewed data when a large population is excluded due to the digital divide \cite{schintler2017constantly}. The digital divide could also significantly affect the policy-making and evaluation process as the feedback of a large population may not be available due to no or limited access to social media and other relevant technologies. Moreover, the digital divide could also significantly affect the impact of social awareness campaigns and contact tracing for various applications. Similarly, due to less exposure to modern technologies, individuals and groups could fall into propaganda and fake news \cite{burkhardt2017combating}.

\subsection{Technical Challenges}
Besides societal challenges, social media informatics applications in sustainable cities and societies also face several technical challenges. In this section, we discuss these challenges in more detail.
\begin{table*}[]
\centering
\caption{Summary of some key challenges, risks, and issues associated with social media informatics for sustainable cities and societies.}
\label{tab:keychallenges}
\scalebox{.7}{
\begin{tabular}{|p{2.5 cm}|p{9 cm}|p{9 cm}|}
\hline
\textbf{Aspects/Concerns} & \textbf{Challenges} & \textbf{Potential Solutions} \\ \hline
Ethical and Privacy Concerns & 
 Informed Consent; Data Bias; Transparency and accountability & It requires a multifaceted approach by involving different stakeholders, such as ethicists, legal experts, and
social scientists, and maintaining a balance between the benefits of social media informatics and the protection of individual rights. \\ \hline
Data Ownership & Deciding who should have control over the data? The public or the businesses who invested in the infrastructure/application; A more control of 
public over the data would harm
the investors' interests while control of business may compromise public privacy and rights?  &  The public authorities/administration could seek a fine balance between the business and citizens’ control over the data could help; Similarly, the newly introduced concept of nationalization of data could help where public authorities could have control
over the data.\cite{ahmad2022developing}. \\ \hline
Public Trust and Perception & the growing privacy concerns due to several data breaches
incidents; the abundance of propaganda, misinformation, and fake news; extensive data manipulation through data moderation & A multifaceted approach is required to build and maintain public trust in social media informatics through transparency, accountability, ethical practices,
public engagement, and effective communication. \\ \hline
Legal and Regulatory Issues &  The legal and regulatory concerns, such as transparency \& accountability, safety \&
surveillance and privacy \& confidentiality make processing social media data more challenging.  &  A fine balance between social media data usage and legal \& regulatory laws implementation could improve the situation. \\ \hline
Misinformation and Fake News& Lack of
central control, manipulation of information, and lack of proper verification mechanisms.   &   A multifaceted approach involving social media users in the verification process and AI-based solutions could improve the accuracy and authenticity of information shared on social media. Moreover, policies and regulations
could also help.\\ \hline
Echo Chambers&  Spreading propaganda and misinformation, reinforcing certain beliefs and opinions, influencing public views for political, religious, business, and other motives,
and censorship and control over social media content. & A multi-faceted approach is required by engaging all stakeholders and deploying state-of-the-art social media informatics techniques and user behavior changes. \\ \hline
Digital Divide& Bias in decision-making due to AI algorithms trained on skewed
data when a large population is excluded due to the digital divide. It also significantly affects
the policy-making and evaluation process as the feedback of a large population may not be available.  Moreover, with less exposure to modern technologies, individuals and groups could fall into propaganda and fake news & Investment in infrastructure and equal opportunities to all groups and communities; Investment in technological education especially adult lifelong learning.\\ \hline
Data Crawling/Scraping & Social media platforms have different data access restrictions and API limitations, such as rate limits. APIs are also subject to changes, involving the removal/depreciation of certain
features, that could affect the data scraping process.  & Different strategies, such as collecting data from multiple sources/social
media platforms, data prioritization based on the relevance to the application, the use of multiple and premium (with higher
data limits) accounts for APIs and the use of web scraping techniques could help. \\ \hline
Data Pre-processing & Abundance of unnecessary information; The correction of OOV words introduced due to intentional and unintentional textual variations, which are generally to bypass different filters for different applications, such as hate speech, propaganda, and sarcasm
detection. &  More sophisticated pre-processing methods are required to deal with
application-specific challenges, consider the context of OOV words, and pre-processing techniques that can be adapted to different languages and domains. \\ \hline
Adversarial Attacks &  Adversarial attacks could be
launched to fool key sustainable cities and society applications, such as fake news, hate speech detection, and bias detection,
to hide malicious activities on social media. Adversarial attacks could also be designed to provoke or intensify enraged mobs or create social division by promoting polarizing content and targeting vulnerable communities. Moreover,
adversarial attacks could be used to mislead policymakers by presenting manipulated public opinions. & Handling OOV words during pre-processing could remove adversarial attacks based on textual variations; Explainability could also help develop robust social media informatics algorithms.\\ \hline
Dynamic and Evolving Nature of Social Media & Social media platforms and user behaviors constantly evolve, maintaining up-to-date and relevant analytics more challenging.  & adaptive
multifaceted approaches involving different stakeholders are required. The data collection methods, policies, and methodologies need to be updated by flexible to changes. Similarly, social
media informatics frameworks, algorithms, and AI/ML models need to be continuously updated to keep pace with these
changes.
 \\ \hline
\end{tabular}}
\end{table*}

\subsubsection{Data Crawling/Scraping}
Data crawling/scraping is the first step and most crucial part of any social media informatics application. The success of any social media informatics application largely depends on the availability of sufficient and quality data \cite{ahmad2022developing}. Several factors, such as user privacy, load management of the platforms' servers, and preventing misuse of the data, are involved with users' data on social media platforms, making data scraping more challenging. For instance, these social media platforms have different data access restrictions and API limitations. These APIs have a rate limit allowing a certain number of requests per unit time and the crawlers have to wait once the number of requests exceeds the limit. Similarly, social media platforms provide various levels of access to APIs with different access rights in terms of the amount and type of current and historical data to be scraped.  Similarly, these APIs are subject to changes, involving the removal/depreciation of certain features, that could affect the data scraping process. Different strategies, such as collecting data from multiple sources/social media platforms, data prioritization based on the relevance to the application, the use of multiple and preimum (with higher data limits) accounts for APIs, and the use of web scraping techniques (by following the terms and conditions of the platforms) \footnote{What Is Web Scraping?: https://careerfoundry.com/en/blog/data-analytics/web-scraping-guide/}, could be used to overcome these challenges.  

\subsubsection{Data Pre-processing}
Social media data is generally noisy, unstructured, and inconsistent, which makes it difficult to process. Therefore, data is cleaned and prepared for processing using various strategies and techniques during the data processing step. Data pre-processing is one of the key components of social media informatics frameworks and involves several tasks, such as data cleaning, tokenization, stemming, lemmatization, normalization, data balancing, and handling of out-of-vocabulary (OOV) words. In data cleaning, unnecessary information, such as URLs, short words, punctuation marks, numbers, emojis, special characters, and white spaces, is removed. However, in certain applications, such as sentiment analysis author style detection \cite{zamir2023document}, these special characters, emojis, and punctuation marks could be useful. Text normalization also involves several key tasks, such as expanding contracts into their full form, handling dialectal words, removing repetition of alphabets, and phonetic substitution \cite{khan2021enhancement}. Similarly, social media content is subject to intentional and unintentional textual variations, which results in OOV words. The intentional variations in the text are carried out to bypass different filters for different applications, such as hate speech, propaganda, and sarcasm detection. To improve to effectiveness of social media informatics algorithms, these intentional and unintentional textual variations must be handled during reprocessing \cite{khan2020enhancement}.  

The literature reports several interesting solutions to deal with challenges associated with noisy and unstructured data, data balancing, text normalization, and handling OOV words. However, considering the dynamic nature of social media data, the traditional reprocessing methods are not enough to address these challenges \cite{khan2021enhancement}. To address these challenges more sophisticated preprocessing methods are required to deal with application-specific challenges, consider the context of OOV words, and pre-processing techniques that can be adapted to different languages and domains.  

\subsubsection{Computational Challenges}
Another key technical challenge associated with social media informatics is the requirement for extensive computational resources. Social media informatics applications for sustainable societies generally involve processing and analyzing large volumes of multimodal data (text, images, videos, and audio clips), which requires extensive computational resources \cite{sapountzi2018social}. Similarly, various social media informatics applications for sustainable cities and societies involve real-time processing of multi-modal data for tasks, such as event detection for disaster management \cite{ahmad2019social}, requiring extensive computational resources. Similarly, some of the applications, such as contact tracing and predictive policing for crime control, involve Graph processing and analysis of large graphs representing users, their connections, and interactions. Graph analysis algorithms are computationally intensive especially when large networks are involved \cite{sandryhaila2014big}. To overcome these computational challenges, social media informatics frameworks require more robust data collection, management, and processing strategies, efficient algorithms, and training approaches.

\subsubsection{Adversarial Attacks}
Similar to other ML and AI applications, adversarial attacks pose significant threats to social media informatics applications for sustainable societies and cities \cite{alsmadi2022adversarial}. Several factors attract malicious users/attackers to target social media informatics solutions for sustainable cities and societies. For instance, the attacks could be launched to manipulate information and public sentiments \cite{de2021adversarial}.  Similarly, adversarial attacks could be launched to fool key sustainable cities and society applications, such as fake news, hate speech detection, and bias detection, to hide malicious activities on social media \cite{koenders2021vulnerable}. Similarly, adversarial attacks could be designed to provoke or intensify enraged mobs or create social division by promoting polarizing content and targeting vulnerable communities. Moreover, adversarial attacks could be used to mislead policymakers by presenting manipulated public opinions leading to policies that may not align with sustainability goals \cite{alloulbi2022retracted}. Predictive policing is also one of the most vulnerable applications to adversarial attacks, where the adversaries disturb the predictive capabilities of the AI/ML model \cite{helm2021beyond}. Similarly, these adversarial attacks could also be used to fool and bypass the intrusion detection algorithms for the security of sustainable cities and society's infrastructure \cite{khan2024explainable}. 

The adversaries generally rely on textual variations to disturb the predictive capabilities of social informatics models. For instance, Khan et al. \cite{khan2023efficient} utilized textual variations in social media content for crafting efficient character-level adversarial attacks for different social media informatics applications. Thanks to the recent development in adversarial ML, efficient strategies and techniques have been proposed to guard against such adversarial attacks, leading to the development of robust sustainable cities and societies' applications \cite{chen2021cyber,ahmad2022developing}. One of the potential research directions in this regard is the potential of explainability in coping with adversarial attacks \cite{ahmad2022developing}. However, the use of explainability in adversarial ML is a double-edged sword as the explanations provided by explainability enable the attackers to launch more efficient attacks on social informatics frameworks for sustainable cities and societies \cite{khan2024explainable}. 

\subsubsection{Dynamic and Evolving Nature of Social Media }
Another challenge with social media informatics for sustainable cities and societies is adapting to the new trends in social media. Social media platforms and user behaviors constantly evolve, maintaining up-to-date and relevant analytics more challenging. For instance, the recently introduced Internet of Behaviors (IoBs) opens new opportunities and challenges in social media informatics \cite{patel2022internet}. Some of the common challenges introduced to social media informatics applications for sustainable cities and societies due to the evolving nature of social media include an enormous increase in the amount of data to process as a result of new features by social media platforms, ethical and privacy concerns, technical challenges in terms of misinformation and adaptive models and approaches to new events, and an increase in the cybersecurity threats and adversarial attacks. To address these challenges, adaptive multifaceted approaches involving different stakeholders are required. The data collection methods, policies, and methodologies need to be updated by being flexible to changes. Similarly, social media informatics frameworks, algorithms, and AI/ML models must be continuously updated to keep pace with these changes.

\section{Insights and Lessons Learned}
\label{sec:insights}
This section summarizes the key insights and lessons from the literature on social media informatics for sustainable cities and societies.

\begin{itemize}
    \item Over the last few years, social media emerged as a useful source of information for several interesting application domains.
    \item Similar to other application domains, social media informatics has a great potential to significantly contribute towards more sustainable cities and societies by enabling various smart cities and societal applications. 
    \item The literature reports several interesting studies exploring different aspects of social media informatics for sustainable cities and societies applications.
    \item The number of publications produced per year has increased year by year with very high growth in the current year. This shows the interest of the community in the domain and the potential of social media informatics in smart cities and sustainable societies.
    \item Despite several successful deployments, social media informatics for sustainable cities and societies applications faces several technical, ethical, and societal challenges, making deploying these frameworks very challenging. 
    \item These challenges significantly limit its effectiveness in achieving sustainable cities and societies and need a joint and multifaceted approach involving all the stakeholders to tackle and overcome these challenges.
    \item On one side, the technical aspects need to be considered by refining the data collection methods, policies, and methodologies. Similarly, social media informatics frameworks, algorithms, and AI/ML models must be updated continuously to keep pace with new trends and challenges.
    \item On the other side, policies, and regulations must be implemented for transparency and accountability, allowing appropriate actions against the individuals, organizations, communities, and groups involved in fraudulent, unethical, and wrong practices.
    \item The growing demand for legislation and regulation could have negative consequences. For example, it could change how social media platforms operate and manage users' data, hurting social media informatics for sustainable cities and societies.  
    \item To ensure users' privacy \& rights and effective use of social media informatics in sustainable cities and societies, a fine balance between social media data usage and legal and regulatory laws implementation is required.
\end{itemize}


\section{Conclusion}
\label{sec:conclusion}
In this paper, we have analyzed and examined the contributions of social media informatics to sustainable cities and societies by providing a detailed overview of its key applications. We have also analyzed and reported key technical challenges and societal concerns associated with social media informatics for sustainable cities and societies. 
The paper also discussed the potential solutions to these challenges and explored how social media informatics could be made more effective for sustainable cities and societies. This detailed overview of the social media informatics applications in sustainable cities and societies is expected to provide a baseline for future research.

\section*{Acknowledgments}
This research was supported by the Institute of Information \& Communications Technology Planning \& Evaluation (IITP), funded by the Ministry of Science and ICT of the Korea Government (MSIT) under the Artificial Intelligence Convergence Innovation Human Resources Development(IITP-2024-RS-2023-00255968) grant and under Grant RS-2021-II212068 (Artificial Intelligence Innovation Hub), and also by the National Research Foundation of Korea (NRF) grant funded by the Korea government (MSIT) (NRF-2022R1A2C1007434) and the BK21 FOUR program funded by the Ministry of Education (NRF5199991014091).

\bibliographystyle{ACM-Reference-Format}
\bibliography{sample-base}

\end{document}